\documentclass[journal=jacsat,manuscript=article]{achemso}
\usepackage{chemformula} 
\usepackage[T1]{fontenc} 
\usepackage{amsmath} 
\usepackage {units}
\usepackage{here}    
\usepackage{amsmath,amssymb}
\usepackage{graphicx}   
\usepackage{verbatim}   
\usepackage{color}      
\usepackage{bm}
\usepackage{subfigure}  
\usepackage{hyperref}   
\usepackage[normalem]{ulem} 
\usepackage{siunitx} 

\author{Shuzo Kato}
\affiliation[kyushu]{Department of Physics, Kyushu University, Motooka 744, Fukuoka 819-0395, Japan}
\author{David Garenne}
\affiliation[minnesota]{School of Physics and Astronomy, University of Minnesota, 115 Union street se Minneapolis, MN 55455, USA}
\author{Vincent Noireaux}
\affiliation[minnesota]{School of Physics and Astronomy, University of Minnesota, 115 Union street se Minneapolis, MN 55455, USA}
\author{Yusuke T. Maeda}
\affiliation[kyushu]{Department of Physics, Kyushu University, Motooka 744, Fukuoka 819-0395, Japan}
\email{ymaeda@phys.kyushu-u.ac.jp}

\title{Phase separation and protein partitioning in compartmentalized cell-free expression reactions}


\begin{document}

\begin{abstract}
Liquid-liquid phase separation (LLPS) is important to control a wide range of reactions from gene expression to protein degradation in a cell-sized space. To bring a better understanding of the compatibility of such phase-separated structures with protein synthesis, we study emergent LLPS in a cell-free transcription-translation (TXTL) reaction. When the TXTL reaction composed of many proteins is concentrated, the uniformly mixed state becomes unstable, and membrane-less phases form spontaneously. This LLPS droplet formation is induced when the TXTL reaction is enclosed in water-in-oil emulsion droplets, in which water evaporates from the surface. As the emulsion droplets shrink, smaller LLPS droplets appear inside the emulsion droplets and coalesce into large phase-separated domains that partition the localization of synthesized reporter proteins. The presence of PEG in the TXTL reaction is important not only for versatile cell-free protein synthesis but also for the formation of two large domains capable of protein partitioning. Our results may shed light on the dynamic interplay of LLPS formation and cell-free protein synthesis toward the construction of synthetic organelles.
\end{abstract}

\section{Introduction}

The intracellular cytosol of living cells densely encloses proteins and nucleic acid macromolecules. In recent years, it has become evident that membrane-less droplet-like organelles are formed in cells \cite{hymann,brangwynne1}. Such membrane-less structures are involved in various cellular functions, including transcription control and genome organizations \cite{mekhail}. The underlying process that drives the formation of these droplet-like bodies is liquid-liquid phase separation (LLPS). Understanding such intracellular phase separation is also gathering attention in the fields of biological soft matter \cite{falahati,klosin,martin} and droplet-based engineering of RNA/DNA \cite{navarro,liedl,mann}. Recently, it has been shown that the level of oversaturation strongly depends on components in intracellular LLPS \cite{brangwynne2}. In phase separation of complex protein solutions, the characteristics of individual molecules such as hydrophobicity and intrinsically disordered regions result in different molecular partitions. Thus, conventional \textit{in vitro} models of simplified two-phase separation \cite{chao} must be extended for exploring intracellular phase separation. Synthetic cell systems \cite{noireaux1} are convenient experimental platforms for examining the fundamental principles of intracellular LLPS in a simplified context, also including a cytoplasmic extract that recapitulates gene expression.

Transcription-translation (TXTL) reactions are emerging as versatile tools for constructing and interrogating biological systems from the ground up \cite{jewett,noireaux2,noireaux3, macdonald,hibi}. It can be achieved in a broad range of physical settings, from test tubes to cell-sized compartments referred to as synthetic cells \cite{lu,yomo1,kuruma2,simmel}. Cell-sized TXTL reactors have been constructed in lipid vesicles \cite{noireaux2,noireaux3}, emulsions \cite{sakamoto}, and microfluidic chips \cite{barziv,ziane1}. In these reactors, an enclosed TXTL reaction autonomously expresses genes or gene circuits. These technical settings have enabled investigating the design principles of information processing for cell-cell communication \cite{barziv2} and the self-organization of cytoskeletons \cite{maeda1,kuruma,noireaux4}. TXTL reactions can also be achieved in membrane-less droplets \cite{tang,wilhelm1,wilhelm2,mansy}. Such droplets crowded nature provides the advantage of enhancing transcriptional activity \cite{leduc,wilhelm3}. To this end, cell-free expression carried out in membrane-less compartments created through LLPS is a practical tool for understanding the physical properties (e.g., molecular sorting and trapping) and biological functions (e.g., gene expression) of the phase-separated structures of complex biological fluids. Because the construction of cell-sized compartments that host TXTL reactions is at an early stage, many biophysical and biochemical properties remain to be discovered in these synthetic cell systems, which are gaining increasing attention.

This study describes a synthetic cell system capable of cell-free gene expression and intracellular phase separation. We enclosed a TXTL reaction prepared from \textit{Escherichia coli} (\textit{E. coli}) \cite{noireaux4,noireaux5} in actively shrinking water-in-oil (W/O) emulsion droplets \cite{izri,edel,collier,kopp}. As the volume fraction of the TXTL reaction increased, uniform solubilization became unstable and LLPS is observed. When two different reporter proteins were synthesized simultaneously in the TXTL reaction, each synthesized protein was separately accumulated and sorted into different compartments, demonstrating that the TXTL reaction stays active when LLPS was achieved. The synthetic cells used in this study are thus a convenient experimental model to reconstruct phase separations in a reaction that mimics the intracellular environment and provides new insights into how a spatial structure is created.

\section{Experimental section}
\subsection{Chemical reagents}
The reaction mix contained a cell-free system (myTXTL, Arbor Bioscience) and 2.0\% polyethylene glycol (PEG, molecular weight 8000)\cite{noireaux6,noireaux8}. Plasmid DNA (PLtetO1-mCherry, final concentration of 1.36 nM and/or PLtetO1-deGFP, final concentration of 1.39 nM) was added to the mixture for cell-free gene expression. The marker proteins, the purified deGFP protein (final concentration of \SI{0.49}{\micro M}) and the purified mCherry protein (final concentration of \SI{0.43}{\micro M}), are also used to test protein localizations. The volume ratio of the reaction mix was 9 (extracts):0.5 (marker protein):1.5 (plasmid DNA):1 (d$\rm{H_2O}$), and plasmid DNA was replaced with d$\rm{H_2O}$ or marker proteins for the control experiment. The list of the plasmids, marker proteins, and the reaction mix is shown in Table S1.

\subsection{Preparation of synthetic cells}
The synthetic cell was fabricated using ($\Delta$9-Cis) Phosphatidylcholine (18:1, Avanti Polar Lipids) as a surfactant, which was dissolved in mineral oil (Light mineral oil M5904, Sigma-Aldrich) at 0.1\%w/v. The reaction mix with a volume of \SI{12}{\micro\liter} was added to the oil and emulsified by tapping the tube. The W/O emulsion containing the TXTL reaction is formed in a lipid monolayer of \SI{10}{\micro\meter} - \SI{100}{\micro\meter} in size, which is defined as a synthetic cell. For microscopic analysis, synthetic cells were placed on a solid substrate covered with poly-dimethyl siloxane (PDMS, Dow corning Sylgard184) within a square-frame sealed chamber (SLF0601, Bio-Rad) and then closed with a PDMS block.

\subsection{Optical microscopy and image analysis}
All images, including time-lapse images, were obtained using a confocal microscope (IX83 inverted microscope from Olympus, CSU-X1 confocal scanning unit from Yokogawa Electric Co. Ltd., and iXon-Ultra EM-CCD camera from Andor Technologies). All microscopic observations were obtained at an exposure time of \SI{100}{\milli\second} and the temperature was kept at 30$^\circ $C using a homemade copper chamber on the microscope stage. Because LLPSpreferentially occurs at lower temperatures, the temperature for gene expression was set at 30$^\circ $C to reduce the temperature to the maximum possible extent while maintaining the gene expression activity.
In Figures 2 and 3, the fluorescence intensity of purified deGFP or mCherry was used to estimate the number of protein molecules. By measuring the fluorescent intensity of a marker protein in the closed chamber, we calculated the total fluorescence intensities of a unit volume and plotted it as a function of the number of protein molecules in fmol (Figure S1). From the obtained calibration curve, the fluorescent intensity of the reporter protein was converted to fmol for all data. Image analysis was performed using the Image Processing Toolbox and Computer Vision toolbox of the MATLAB software (MathWorks).

\subsection{Concentration factor of the lysate}
The concentration factor of the lysate $\alpha_v$ is the increased rate of the lysate concentration after dehydration by membrane dialysis. For the measurement of this concentration factor, an epi-fluorescence microscope was used to measure $\alpha_v$ in the concentrated TXTL reaction containing the deGFP volume marker with initial concentration $c_0$ in bulk. The fluorescence intensity of the reaction mix (\SI{96}{\micro\liter}) containing purified GFP, $I_0$, was measured before the centrifugation. We assumed that the number of deGFP volume markers was constant after centrifugation and then recorded the average intensity of deGFP in the condensed reaction. We measured the deGFP fluorescence of this concentrated lysate, $I$, to estimate the concentration factor $\alpha_v = \frac{I}{I_0}$.

\section{Results and discussion}

\subsection{LLPS of the lysate for TXTL reactions}
The lysate used for TXTL reactions is a cytoplasmic extract that contains 9-\SI{10}{\milli\gram\per\milli\liter} of soluble proteins from the \textit{E. coli} cytoplasm. Its protein composition was determined by mass spectrometry recently \cite{noireaux6}. This lysate is under physiological conditions: 150-200 mM salts and pH 7.5-8.0. It is known that LLPS occurs owing to the entropic effect of polymers and water solvent in a high-concentration polymer solution, for example, dehydration effect \cite{park}. First, we performed the condensation of the lysate to examine whether phase separation is induced as the molecular concentration increases. 

\begin{figure}[tb]
\centering
\includegraphics[scale=0.45,bb=0 0 552 592]{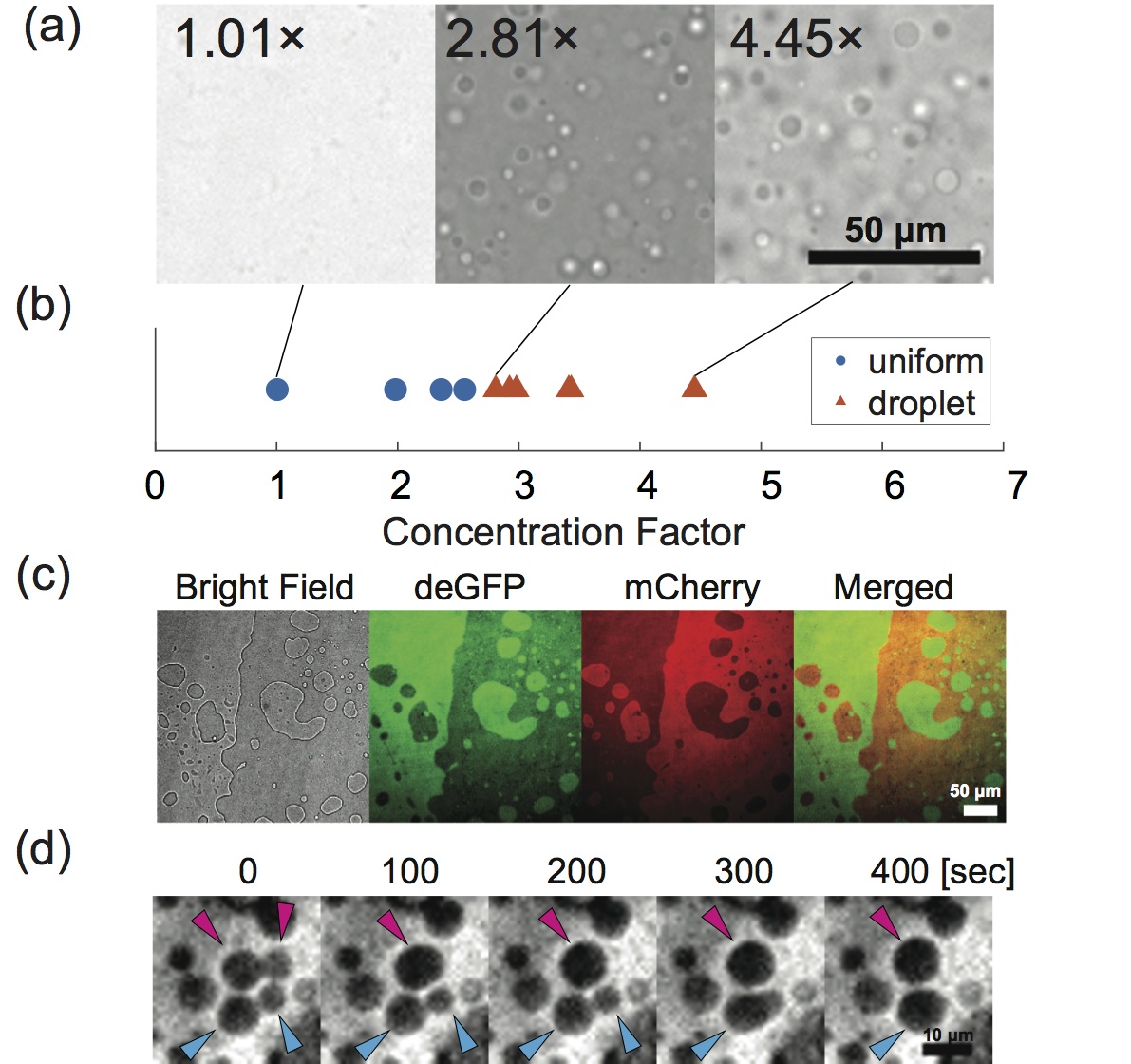}
\caption{\textbf{LLPS of the TXTL reaction in a closed chamber}. (a) The TXTL reaction at different concentration factors: $\alpha_v = 1.00$ (left), $\alpha_v = 2.81$ (center), and $\alpha_v = 4.45$ (right). Scale bar: \SI{50}{\micro\meter}.
(b) Diagram of LLPS observed in the condensed TXTL reaction as a function of $\alpha_v$. Droplets are observed in regions where $\alpha_v$ is higher than 2.81.
(c) Fluorescence images of phase separation in condensed TXTL reactions. Scale bar: \SI{50}{\micro\meter}. Purified deGFP (Green) and purified mCherry (Red) form segregated domains in the condensed TXTL reaction ($\alpha_v = 3.71$). 
(d) Fluorescence images of droplets observed at $\alpha_v = 3.43$. The region in white color represents the deGFP-rich compartment. Droplets coalesce to form a large droplet (unit: second). The droplets indicated by the light blue and magenta arrows exhibit coalescence when they come in close contact. Scale bar, \SI{10}{\micro\meter}. The height of the chamber was \SI{10}{\micro\meter} (Figure \ref{fig1}c) and \SI{300}{\micro\meter} (Figure \ref{fig1}d). Figure \ref{fig1}c, where the phase-separated domain was confined to a thin chamber, shows the irregular boundary due to friction with the PDMS substrate.}\label{fig1}
\end{figure}

As a measure of the relative increase in the volume fraction, the rate of increase in volume was defined as the concentration factor $\alpha_v = \frac{V_0}{V}$, where $V_0$ is the initial volume of the lysate for TXTL reactions, and $V$ is the final volume after centrifugation. $\alpha_v$ is referred to as the concentration factor. Then, we prepared a concentrated lysate with $\alpha_v$ ranging from 1.00 to 4.45 using high-speed centrifugation and a filter device of molecular mass cutoff of 3 kDa. When $\alpha_v$ was less than 2.81, the lysate solution was transparent without any aggregation bodies (Figures \ref{fig1}a and \ref{fig1}b). Intriguingly, when $\alpha_v$ was larger than 2.81, the lysate's transparency abruptly changed, and droplet-like bodies were observed. These bodies were spherical, and the radius ranged from \SI{1.0}{\micro\meter} to \SI{25}{\micro\meter} (Figure \ref{fig1}a). The bodies were stable even when $\alpha_v$ increased to 4.45. The uniform solubilization becomes unstable when the concentration factor exceeds the transition threshold close to $\alpha_v =2.81$, which corresponds to \SI{28} {\milli\gram\per\milli\liter} as a critical concentration of soluble proteins.

\begin{figure*}[tbp]
\centering
\includegraphics[scale=0.54,bb=0 0 632 702]{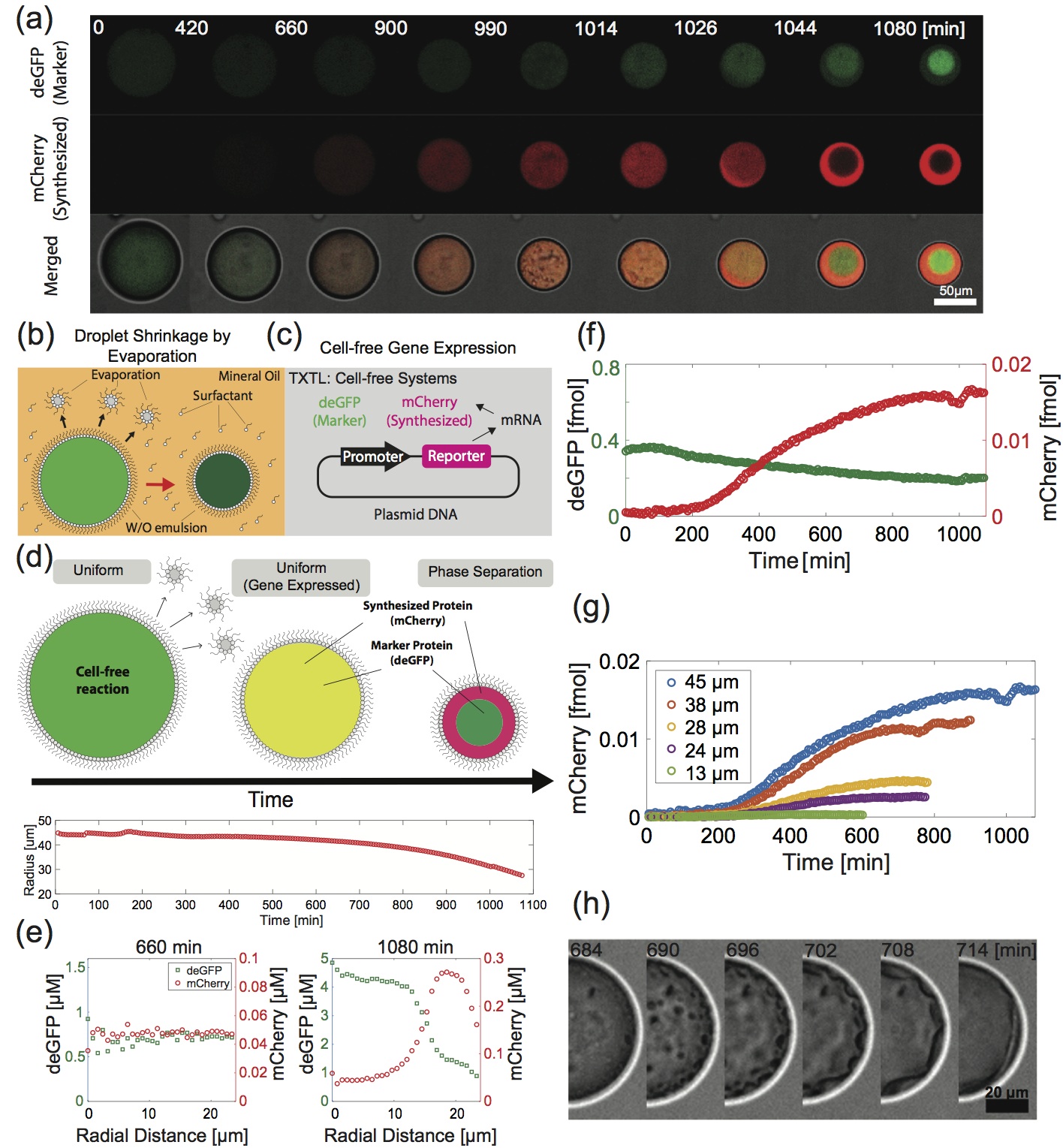}
\caption{\textbf{LLPS of the TXTL reaction in a cell-sized emulsion droplet}. (a) Time series of confocal microscopy images of an emulsion droplet containing a cell-free reaction and plasmid DNA to dynamically synthesize mCherry protein over time. The initial radius of the emulsion droplet is $R_0 = \SI{45}{\micro\meter}$. The purified deGFP marker protein was added. deGFP and mCherry are phase separated at critical volume compression. Scale bar, \SI{50}{\micro\meter}.
(b) Schematic of active interface due to evaporation of confined water. This dehydration process reduces the fraction of water solvent inside emulsions and then induces the reduction in emulsion radius.
(c) Genetic circuit encoded in plasmid DNA. \textit{mCherry} gene is expressed through the PLtetO1 promoter. The expression of \textit{mCherry} gene starts in the TXTL reaction inside the emulsion droplet.
(d) Schematic of the phase transition of the shrinking emulsion (upper panel). The representative shrinkage dynamics of emulsion radius is shown in the bottom panel.
(e) Cross-sectional fluorescent intensity profiles of purified deGFP marker protein (green) and synthesized mCherry protein (red). In the early phase at $t = \SI{660}{\minute}$ (left), the fluorescence intensity is uniform for both proteins. After the phase transition at $t = \SI{1080}{\minute}$ (right), the spatial distributions of both proteins do not overlap with each other, which indicates phase separation, 
(f) Kinetics of synthesized mCherry protein from gene expression (red) and purified deGFP marker protein (green). 
(g) Kinetics of \textit{mCherry} gene expression in different-sized emulsions.
(h) Formation of small droplets and their coalescence close to the transition point of LLPS. We note that the droplet shown in Figure \ref{fig2}h is different from the one shown in Figure \ref{fig2}a. Scale bar, \SI{20}{\micro\meter}.
}\label{fig2}
\end{figure*}

One of the significant properties of LLPS is membrane-less compartmentalization, in which proteins, DNA, and other polymers and electrolytes can be accumulated in a separate domain and distributed to compartments. To test whether the droplet-like bodies in the concentrated TXTL reaction induced protein partitioning, we added two different purified fluorescent proteins, deGFP (a final concentration of \SI{5.22}{\micro M}) and mCherry (\SI{4.34}{\micro M}), to the concentrated reaction. We measured the spatial distributions of deGFP and mCherry using a confocal fluorescent microscope. The spatial distribution of the two reporter proteins was uniform when $\alpha_v$ was less than 2.81 (data not shown). As $\alpha_v$ increased beyond the threshold and LLPS occurred, for example, $\alpha_v = 3.71$ in Figure \ref{fig1}c, the two proteins spontaneously accumulated in different phase-separated regions created by LLPS (Green: deGFP and Red: mCherry) (Figure \ref{fig1}c, merged). The two reporter proteins were sorted out in two different membrane-less compartments.

We conducted the time-lapse imaging of the droplet-like bodies at $\alpha_v =3.43$ closer to the transition point to ensure their spontaneous coalescence. The dark region (corresponding to the mCherry-rich phase of Figure \ref{fig1}c) showed the coalescence of two smaller bodies (diameter of \SI{2.5}{\micro\meter}). It subsequently relaxed into a spherical shape with a larger size (diameter of \SI{4.5}{\micro\meter})(Figure \ref{fig1}d, Movie S1). The observed droplet coalescence is consistent with the fundamental nature of an LLPS droplet without a lipid membrane. In addition, the relaxation into a spherical shape suggests the presence of an interface maintained by surface tension.

\subsection{Active interface induces LLPS and compartmentalization in synthetic cells}

LLPS droplets accumulate and sort proteins into membrane-less compartments \cite{albertson}. The sorting of purified proteins has been demonstrated in membrane-less domains inside a liposome \cite{rivas,dekker}. We aimed to examine how LLPS can be induced to target the spontaneous compartmentalization of synthesized proteins while performing cell-free gene expression. 
However, we found that twofold concentration of the lysate reduced gene expression activity to less than one-tenth, even though the concentration factor, $\alpha_v = 2.07$, was such that LLPS did not occur (Figure S2) \cite{nomura}. This reduced gene expression capacity indicates that the TXTL reaction was no longer optimal after rapid forced condensation. A physical process was needed to maintain gene expression activity with gradual condensation toward the phase-separated state. To this end, we employed a dehydration process during the evaporation of the W/O droplets.

We constructed a synthetic cell system using actively shrinking droplets in a W/O emulsion in which the concentrating factor $\alpha_v$ gradually increases due to dehydration (Figure \ref{fig2}a). It is known that the dehydration of confined water can occur from a lipid monolayer at the surface of a W/O emulsion (Figure \ref{fig2}b) \cite{izri,edel,collier,kopp}. We dissolved a neutral phospholipid (1,2-dioleoyl-sn-glycero-3-phosphocholine) in the oil phase at 0.1\%w/v. The water solvent was gradually depleted by evaporation of the emulsion droplet, which subsequently started to shrink.

To test this design, we encapsulated the TXTL reaction, plasmid DNA to dynamically synthesize mCherry (PLtetO1-mCherry, final concentration of  \SI{1.36}{\nano M}), and purified deGFP marker protein (final concentration of \SI{0.49}{\micro M}) within a W/O emulsion (Figures \ref{fig2}b and \ref{fig2}c). We then measured the fluorescent signals from mCherry and the purified deGFP volume marker using a confocal microscope. The initial radius of the emulsion droplet $R_0$ is \SI{45}{\micro\meter}, as shown in Figure \ref{fig2}a. Its radius decreased monotonically with time (Figure \ref{fig2}d), which confirmed the dehydration process due to the evaporation of confined water. We note that it is necessary to make the substrate glassy in order to suppress this evaporation, but the protein production of the TXTL reaction is drastically reduced in the glass chamber (Figure S5). Therefore, we decided to use a synthetic cell that expresses genes while evaporation occurs.

Figure \ref{fig2}a shows the time-lapse recording of the fluorescent signal of the purified deGFP marker protein (top) and that of the mCherry protein synthesized via cell-free gene expression (middle) in the emulsion (Movie S2). During the early phase up to $t = \SI{660}{\minute}$, the deGFP marker protein and the synthesized mCherry protein were uniformly distributed within the emulsion (Figure \ref{fig2}e, left). The total amount of synthesized mCherry protein inside the single emulsion increased exponentially and then saturated later, whereas the total amount of the deGFP marker protein remained constant (Figure \ref{fig2}f). At a particular time ($t = \SI{990}{\minute}$), the deGFP marker protein and synthesized mCherry protein exhibited a spatially segregated pattern inside the emulsion (top and middle rows in Figure \ref{fig2}a, respectively). Subsequently, at $t = \SI{1026}{\minute}$, the two proteins were accumulated in smaller spherical clusters exclusive of each other. Finally, at $t = \SI{1080}{\minute}$, the synthesized mCherry protein and deGFP marker protein were fractionated in the region beneath the lipid monolayer and at the center of the droplet, respectively (Figure \ref{fig2}e right). We analyzed the concentration ratio of deGFP and mCherry in each phase-separated compartment. The concentration ratio of the marker protein-rich region to the marker protein-depleted region was $5.16 \pm 0.86$ for deGFP and $2.96 \pm 0.60$ for mCherry.

The amount of mCherry protein synthesized by gene expression was observed to increase with the size of the emulsion droplets (Figure \ref{fig2}g). Cell-free gene expression reached a steady state earlier than the LLPS of the reaction under droplet evaporation. The gradual evaporation from the surface is important for protein synthesis without a total loss of the optimized activity.

\begin{figure*}[htbp]
\centering
\includegraphics[scale=0.5,bb=0 0 872 592]{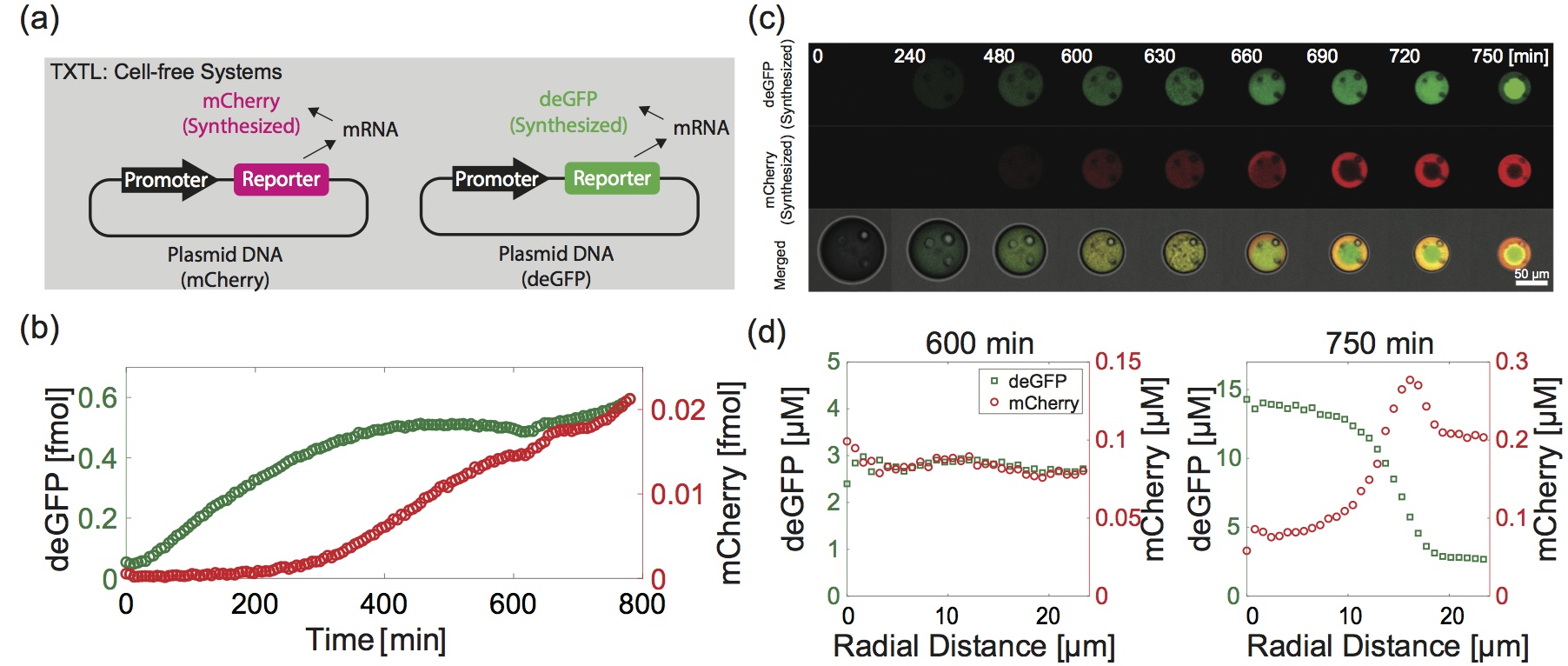}
\caption{\textbf{LLPS and co-expression of genes in a cell-sized emulsion droplet}. (a) Genetic circuit encoded in plasmid DNA. Both \textit{mCherry} gene and \textit{deGFP} gene are expressed through the PLtetO1 promoter. The expression of two gene starts in the TXTL reaction inside the emulsion droplet.
(b) Kinetics of synthesized mCherry protein (red) and synthesized deGFP protein (green) from the gene expression in the emulsion droplet of $R_0 = \SI{47}{\micro\meter}$. The time of LLPS onset is indicated in the dashed line with $T_{PS}$. We note that the kinetics of deGFP synthesis was faster than that of mCherry synthesis (Figure \ref{fig3}b) because maturation of mCherry takes longer time than deGFP in the TXTL reaction \cite{mueller, noireaux5}.
 (c) Time series of confocal microscopy images of a synthetic cell containing a cell-free reaction and plasmid DNA to dynamically synthesize deGFP and mCherry proteins. The spatial distributions of synthesized deGFP (top row) and synthesized mCherry (middle row) are shown in red and green respectively. The initial radius of the emulsion droplet is $R_0 = \SI{47}{\micro\meter}$. Scale bar, \SI{50}{\micro\meter}. (d) Cross-sectional fluorescent intensity profiles of synthesized deGFP protein (green) and synthesized mCherry protein (red). In the early phase at $t = \SI{600}{\minute}$ (left), the fluorescence intensity is uniform for both proteins. After the phase transition (right), two synthesized proteins shows segregated spatial distribution with each other at $t = \SI{750}{\minute}$.
}\label{fig3}
\end{figure*}

The creation of the phase-separated compartments was investigated by observing the detailed internal structure in the emulsion when the phase separation started (Figure \ref{fig2}h). A number of small droplets appeared inside the W/O emulsion. They coalesced to form small droplets that adhered to the oil-water interface. Furthermore, as these droplets coalesced and grew on the emulsion surface, the synthesized mCherry protein was accumulated in a phase-separated compartment. The compartmentalization mediated by coalescence inside the emulsion is consistent with the droplet fusion shown in Figure \ref{fig1}d.

Furthermore, we performed the cell-free synthesis of both deGFP and mCherry proteins in the synthetic cells (Figure \ref{fig3}a and Movie S3) to examine whether the synthesized proteins were accumulated and separated into different regions by LLPS. The emulsion droplet of initial radius $R_0 = \SI{47}{\micro\meter}$ expressed deGFP and mCherry (Figure \ref{fig3}b), and these synthesized proteins uniformly distributed inside the synthetic cell up to $t \approx \SI{600}{\minute}$ (Figure \ref{fig3}c). Before the onset of LLPS, the demixing of the TXTL reaction occurs around at $t = \SI{660}{\minute}$. Then, the two dynamically synthesized proteins were separately localized. We thus concluded that LLPS of the TXTL reaction proceeds inside the synthetic cell while expressing multiple genes and sorting protein products. Notably, the localization of deGFP into the inner domain was not clearly built at $t = 660$ min when phase separation occurred. The fractionation of the synthesized deGFP proceeded slowly for 90 min, and protein segregation of the two reporter proteins was built at $t = 750$ min (Figure 3(c) and Movie S3). The protein synthesis is sustained by the remaining TXTL activity, even though the gene expression capacity is reduced, making it difficult to see a clear LLPS domain.

In addition, the cell-free synthesis of reporter proteins reached approximately half of the steady-state values after $t = \SI{300}{\minute}$ for deGFP and $t = \SI{500}{\minute}$ for mCherry for the emulsion of initial radius with $R_0 = 40$-\SI{50}{\micro\meter}. Phase separation occurred after these characteristic times, and their time difference could be essential to achieve cell-free protein synthesis and its fractionation into phase-separated domains. When the TXTL solution is concentrated twofold ($\alpha_v = 2.07$), although LLPS does not occur at this concentration factor, the gene expression activity is already reduced to less than one-tenth (Figure S2). Furthermore, due to the protein partitioning after LLPS, the lysate proteins that constitute the TXTL reaction can be also fractionated in the space. This may also cause the reduction in gene expression activity at highly concentrated states. Thus, it is advantageous for the enrichment process to proceed slowly, because rapid droplet shrinkage may inhibit the TXTL reaction without sufficient protein synthesis. Although the detailed analysis to determine which proteins are divided into which fractions remains to be addressed in a future work, our finding suggests that the dynamics of droplet evaporation would be a key process to achieve both cell-free gene expression and protein partitioning.

\subsection{LLPS formation is PEG concentration dependent}

\begin{figure*}[tbp]
\centering
\includegraphics[scale=0.55,bb=0 0 822 592]{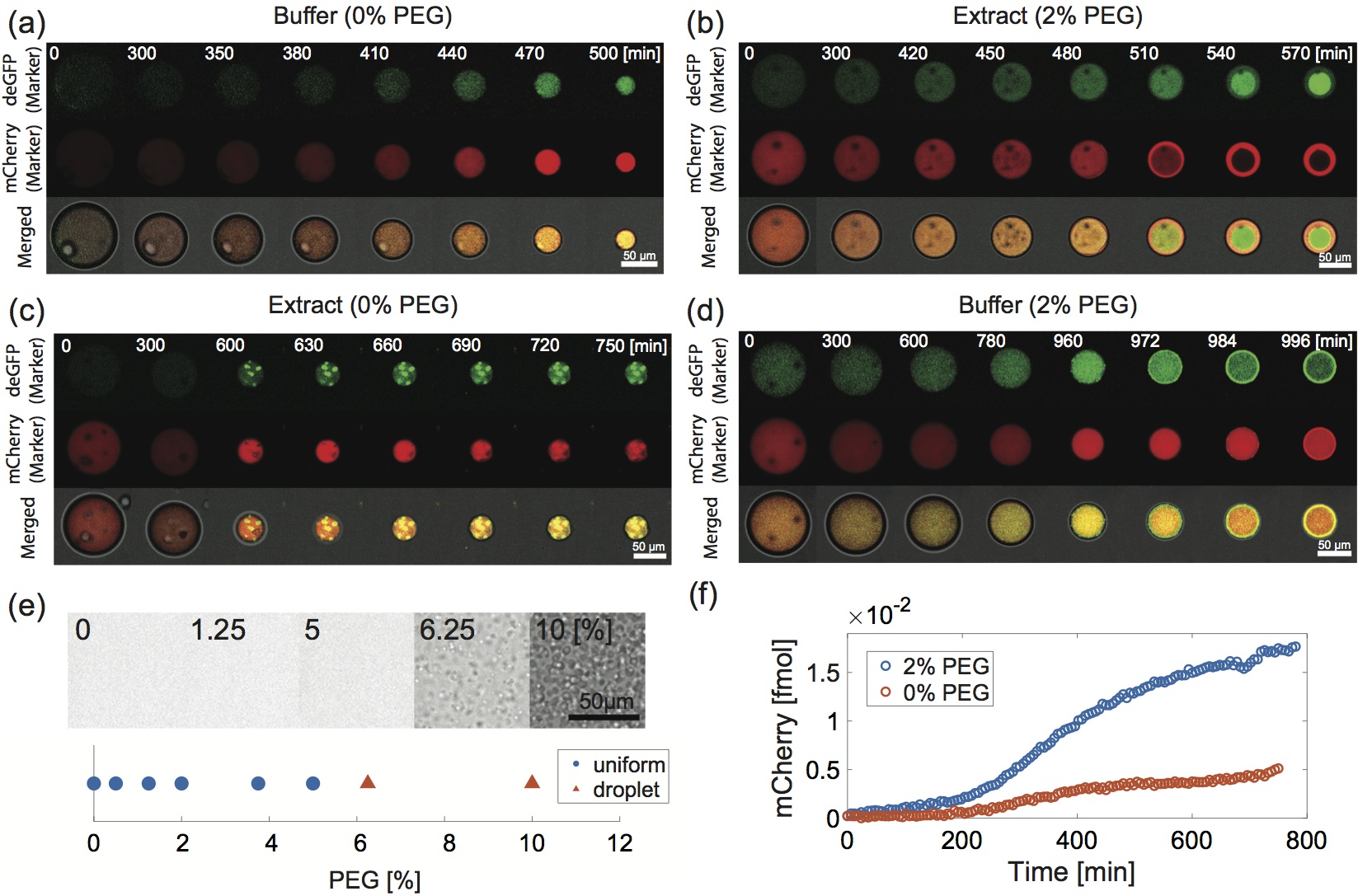}
\caption{\textbf{PEG molecule is essential for the protein partitioning and gene expression activity}. (a)-(d) Top panel: purified deGFP marker and purified mCherry marker, bottom panel: superimposed bright field image. Scale bar, \SI{50}{\micro\meter}. (a) Time evolution of spatial distribution of purified deGFP and mCherry marker proteins in a buffer solution (0\% PEG) without the TXTL reaction. $R_0 = \SI{44}{\micro\meter}$. (b) Time evolution of phase separation in the TXTL reaction with 2.0\% PEG. $R_0 = \SI{46}{\micro\meter}$. (c) Time evolution of spherical aggregate formation in the TXTL reaction without PEG (0\% PEG). $R_0 = \SI{40}{\micro\meter}$. (d) Time evolution of aggregate formation of purified deGFP and mCherry marker proteins in 2.0\% PEG solution without the TXTL reaction. The marker proteins form aggregation patches beneath the membrane interface after 960 min. $R_0 = \SI{46}{\micro\meter}$. (e) Diagram of LLPS observed in the TXTL reaction as a function of the PEG volume fraction. Droplets are observed in regions where the PEG volume fraction is higher than 6.25\% PEG. (f) Kinetics of mCherry expression in the shrinking emulsion with 2.0\% PEG (blue) and 0.0\% PEG (orange).
}\label{fig4}
\end{figure*}

We next analyzed the molecules in TXTL reaction that could be essential for the formation of these large compartments. The TXTL reaction used in this study contained PEG8000 to emulate molecular crowding \cite{noireaux5,noireaux8}. Its volume fraction was approximately 2.0\%, which corresponds to a concentration of 2.5 mM. A previous study has reported\cite{wilhelm1} that PEG is widely used for protein phase separation \textit{in vitro}. We examined the TXTL reaction contents using the purified marker proteins deGFP (final concentrations of \SI{0.49}{\micro M}) and mCherry (\SI{0.43}{\micro M}). As a control experiment, when a solution containing only these two purified marker proteins was enclosed in the emulsions, we observed that LLPS did not occur even after significant condensation (Figure \ref{fig4}a, Movie S4). Next, when the TXTL reaction containing the two marker proteins was confined to W/O emulsions, we found that a two-layered phase-separated structure was formed after shrinkage-induced condensation (Figure \ref{fig4}b, Movie S5), as also seen in the synthetic cells showing gene expression (Figure \ref{fig2} and Figure \ref{fig3}). 

We also examined the LLPS and subsequent protein partitioning of purified deGFP and mCherry marker proteins by enclosing the TXTL reaction without PEG in a W/O emulsion. As shown in Figure \ref{fig4}c, many small spherical bodies containing deGFP appeared as the TXTL reaction's condensation progressed (Movie S6). However, these structures did not grow into a sizeable membrane-free compartment. Next, to investigate the effect of condensation of PEG on protein solubilization and localization, purified marker proteins were enclosed in the emulsion containing 2.0\% PEG alone. Both marker proteins formed aggregates at the oil-water interface; however, unlike LLPS coalescence, the aggregates of deGFP and mCherry did not separately localize inside the emulsion (Figure \ref{fig4}d, Movie S7). The aggregates beneath the membrane could not grow into the larger LLPS domain, implying that the marker proteins in the concentrated PEG solution cannot be solubilized with sufficient hydration water. Our results indicate that TXTL reaction and PEG molecules' coexistence is critical to fractionate proteins into each LLPS domain with higher affinity.

Next, we analyzed whether LLPS of the TXTL reaction would exhibit concentration-dependence on PEG. We kept the concentration of the TXTL reaction constant while varying the PEG concentration from 0\% to 10.0\% to determine the critical PEG concentration where LLPS occurs. We found that the LLPS droplets appeared from the uniformly dissolved solution when the PEG concentration was as high as 6.25\% (Figure \ref{fig4}e). The concentration factor at the onset of LLPS obtained from Figure 1b is 2.81-fold, which indicates that LLPS occurred when the initial 2.0\% PEG was condensed to 5.6\%. The consistency in the critical concentration implies that the coexistence of the reaction components and PEG underlies the formation of the LLPS structure observed in the shrinking emulsion. PEG is necessary not only for the LLPS formation but also to emulate molecular crowding of the TXTL reaction to achieve a significant level of gene expression (Figure \ref{fig4}f). Our experiments thus found that the TXTL reaction with PEG is necessary for both cell-free gene expression and protein partitioning in the synthetic cell. We note that a classical LLPS model in a PEG and dextran mixture \cite{keating} also shows similar concentration dependence on PEG. As the PEG concentration increases, its coexistence with dextran becomes unstable due to a difference in each polymer's affinity toward water molecules.

\begin{figure}[tbp]
\centering
\includegraphics[scale=0.45,bb=0 0 682 592]{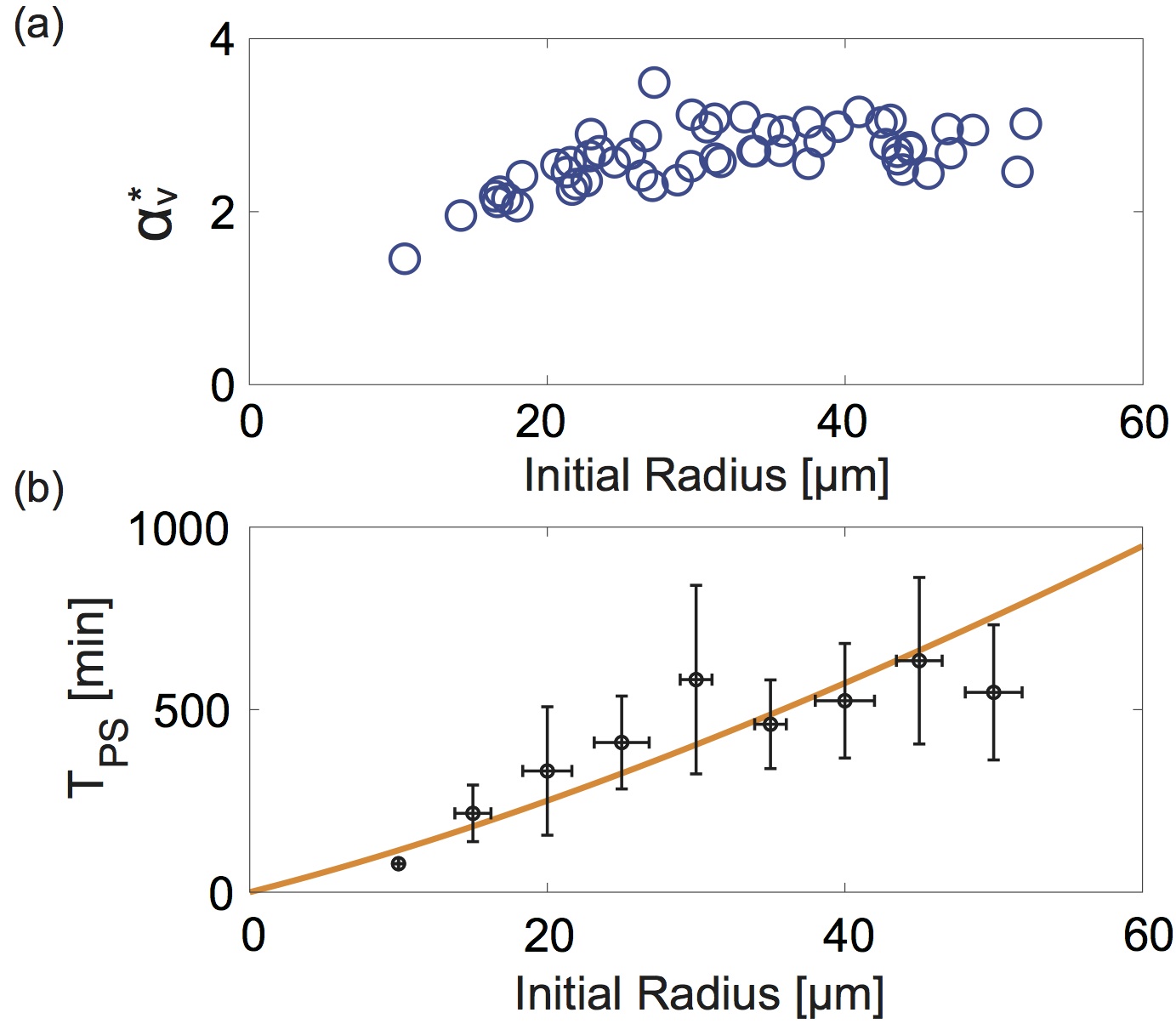}
\caption{\textbf{Size dependence of LLPS in poly-dispersed emulsions}. (a) Scatter plot of concentration factors as function of initial radius of the droplet at the phase separation ($N = 55$ samples). 
(b) Scatter plot of the onset time of phase separation $T_{PS}$ as function of initial radius. $T_{PS}$ is defined by the time when the two proteins of deGFP and mCherry begin to separately localize (Figure S3). The error bars show the standard deviation, and the initial radius on the horizontal axis was averaged every \SI{5}{\micro\meter}. The solid line is the fitting curve of Eq. (1) with $\beta = 2.03(8) \times 10^{-2}$.}\label{fig5}
\end{figure}

\subsection{Surface-to-volume ratio determines LLPS in synthetic cells}

The efficiency of LLPS droplet formation was 100\% (n = 60/60), and protein partitioning during PEG-induced LLPS occurred at the critical concentration of PEG in TXTL reaction. This result further motivated us to examine the critical concentration factor $\alpha_v^{*}$ in shrinking W/O emulsions. We prepared a polydisperse suspension of synthetic cells with different initial volumes, $V_0 = 4\pi R_0^3/3$, under the initial conditions, $t = \SI{0}{\minute}$ and initial radius $R_0$. We measured the dynamics of emulsion shrinkage and the onset of LLPS using time-lapse recording. The onset time of LLPS $T_{PS}$ was defined as the time when the deGFP volume marker and that synthesized mCherry protein started to segregate. We measured the critical radius $R_c$ at $T_{PS}$ to calculate the critical concentration factor as $\alpha_v^{*}=(R_0/R_c)^3$. Figure \ref{fig5}a shows $\alpha_{v}^{*}$ as a function of $R_0$, for the W/O emulsion with $\SI{10}{\micro\meter} < R_0 < \SI{52}{\micro\meter}$. Its mean value $\bar{\alpha_v} = 2.64$ that corresponds to 5.2\% PEG was close to the critical concentration 5.62-6.25\% in bulk (Figures \ref{fig1}b and Figure \ref{fig4}e). We developed a simple model to further examine the almost proportional relationship between $T_{PS}$ and $R_0$ (details are provided in the Supporting Information). We assume that the evaporation of confined water occurs only from the surface. Thus, the rate of volume reduction, $\frac{dV}{dt}$, is proportional to the surface area, $S$. This yields $\frac{dV}{dt} = \beta S$ where $\beta$ represents the shrinkage constant of the emulsion. 

Given that the emulsion is spherical with volume $V = \frac{4 \pi R^3}{3}$ and surface area $S = 4 \pi R^2$, the rate of volume reduction is rewritten as $\frac{dV}{dt} = S \frac{dR}{dt}$. Since a slow decrease in the radius was observed in Figure \ref{fig2}d, we assume that the radius of the droplet, $R$ decays constantly with time, $R(t) = R_0 - \beta t$ ($\beta = 2.03(8) \times 10^{-2}$ obtained from experimental data of emulsion shrinkage). Considering that phase separation occurs when the concentration factor of the TXTL reaction in the emulsion reaches a critical value, $\alpha_v^{*}$, the time at which phase separation occurs is obtained using the following equation: 

\begin{equation}
T_{PS} = \frac{1- (\alpha_v^{*})^{-\frac{1}{3}}}{\beta} R_0.
\end{equation}

The critical concentration factor, $\alpha_v^{*} = 2.02 + 0.0194 R_0$ obtained by the fitting curve from Figure \ref{fig5}a, can be approximated to a constant value because of its weak size dependence (Figure \ref{fig5}a), and $\beta$ is independent of the emulsion size. Thus, Eq. (1) shows the proportional relationship between $R_0$ and $T_{PS}$, which is consistent with Figure \ref{fig5}b.

It is noteworthy that the weak dependence of $\alpha_{v}^{*}$ on $R_0$ implies that large emulsion droplets require a more significant reduction in volume for protein partitioning through LLPS. Figure \ref{fig5}b shows the plot of the onset time of LLPS, $T_{PS}$, and $R_0$ to examine the size dependence of the onset time of LLPS. These parameters exhibited an almost proportional relationship, indicating that the active shrinking interface induces faster LLPS in smaller synthetic cells. The almost proportional relationship can be understood by considering that evaporation is the most effective at a small initial radius because the surface-to-volume ratio is larger in a smaller emulsion droplet.

Notably, the onset of LLPS took more than 10 h for large emulsions, for example, $R_0 >\SI{50}{\micro\meter}$ because the time until phase separation started was determined by the rate of slow dehydration from the W/O interface. One approach to control the dehydration rate is the drifting flow at the oil-air interface layer. The LLPS onset time can be controlled using interfacial drift, and even large droplets can partition gene expression products within a few hours (Figure S6).

\section{Conclusion}
In this study, we demonstrated that LLPS of a TXTL reaction can be achieved by reconcentration. By encapsulating the TXTL reaction into a cell-sized emulsion, cell-free gene expression takes place along with LLPS formation of two large phases capable of segregating proteins in membrane-less compartments. The same observations were made when the two reporter proteins were expressed simultaneously along with the purified reporter proteins. The partitioning of proteins to distinct compartments indicated that slight differences in protein species could be detected by LLPS while keeping the ability of cell-free protein synthesis. Moreover, the phase separation of the TXTL reaction in the shrinking droplet is observed only in the presence of PEG (Figure \ref{fig4}). When fluorescent PEG was added to the reaction, the PEG polymer was localized at the internal LLPS domain (Figure S4). Similar localization of the PEG polymer has been observed in a previous study \cite{wilhelm1}, and the understanding of how proteins interact with a densely concentrated cellular space in the presence of hydrophilic polymers such as PEG can reveal the mechanism of protein partitioning. Furthermore, in the TXTL reaction containing PEG, the phase-separated droplets coalesced to form large domains, whereas these droplets did not fuse and only formed small aggregates in the absence of PEG. This result implies the importance of crowding agents to control the viscoelastic nature of membrane-less compartments. 

Interestingly, the protein concentrations are comparable in bacteria, yeast, and human cells, and their volume fraction of cytosolic proteins is about 20\% in these species\cite{milo}. The typical concentration of total macromolecules (proteins, mRNA, DNA, and PEG) in the TXTL reaction solution that showed LLPS was 10\%. This suggests that the intracellular solution is in the concentration range where both prokaryotes and eukaryotes can sufficiently show LLPS. Phase-separated droplets of RNA polymerase in the bacterial cytoplasm\cite{weber} have been recently reported, which is consistent with our expectation. Besides, for the critical concentration, controlling the size of LLPS droplets is also pivotal to understand the intracellular self-organization. Past study has shown that bacterial carboxysomes form submicron-sized phase-separated droplets \cite{biteen}. In eukaryotic cells, the microtubule associated protein TPX2 forms regularly spaced micron-sized droplets on single microtubules due to surface minimization \cite{petry}. The utilization of physical properties of droplets for controlling the micron-sized droplets is an important step to achieve further spatial structure control in a synthetic cell. 

The synthetic cells generated in this study provide an alternative to conventional models using crowding agents \cite{chao, albertson} for studying LLPS in complex biological environments. This TXTL-based LLPS could be further used to explore other aspects of LLPS, such as the formation of synthetic organelles in liposomes capable of cell-free protein synthesis. The intrinsic surface tension of LLPS droplets may be relevant to induce the shape instability of liposome-based synthetic cells \cite{yuan}. The TXTL reaction-based LLPS may shed light on the dynamic interplay of complex biological reactions and phase separation, which may bring insights into the synthetic organelles \cite{parker,greening,boekhoven} in the liposomes capable of cell-free protein synthesis \cite{noireaux5,ziane1,noireaux8,noireaux7}.

\begin{acknowledgement}

We thank R. Sakamoto and T. Fukuyama for discussion and M. Miyazaki for multipoint measurement in confocal microscopy. This work was supported by Grant-in-Aid for Scientific Research on Innovative Areas JP18H05427 and Scientific Research (B) JP20H01872 (to Y.T.M.), HFSP Research Grant RGP0037/2015 (to V.N. and Y.T.M), and NSF grant EF-1934496 (to V.N.). 

\end{acknowledgement}

\section*{Author contributions statement}

Y.T.M. and S.K. designed research and S.K. conceived the experiments. D.G.  and V.N. prepared the cell-free gene expression systems and the plasmids. S.K analyzed data and prepared the figures. Y.T.M. wrote the manuscript and all authors reviewed the manuscript.

\bibliography{achemso-demo}

\newpage

\setcounter{figure}{0}
\renewcommand{\thefigure}{S\arabic{figure}}
\renewcommand{\figurename}{FIG.}
\renewcommand{\tablename}{TABLE.}
\renewcommand{\thetable}{S\arabic{table}}
\renewcommand{\refname}{References}
\renewcommand{\arraystretch}{1.2}

\section{Supporting Information}

\subsection{The calibration of fluorescent intensity}

We used two proteins, deGFP and mCherry, as reporter proteins. This fluorescent intensity was calculated by measuring the number of proteins from a three-dimensional image of a droplet via the summation of all the fluorescence brightness. The actual amount of protein present was calculated by using a calibration curve between fluorescence intensity and protein amount (Figure S1). We confirmed that the amount of protein synthesized from cell-free gene expression in a single artificial cell was approximately fmol range. The volume occupied by 0.1 fmol of fluorescent protein was estimated to be approximately \SI{0.5}{\micro\meter}$^3$, which corresponds to a volume fraction of less than 10$^{-3}$\% inside a droplet of \SI{40}{\micro\meter} radius; this is negligibly small compared to the proteins in the TXTL reaction. This evaluation indicates that neither the protein synthesized by cell-free gene expression nor the added purified protein affected the appearance of phase separation. By evaluating the total amount of proteins in artificial cells in fmol, we could quantitatively analyze the kinetics and yield of the synthesized proteins.

\begin{figure}[h]
\centering
\includegraphics[scale=0.54,bb=0 0 822 392]{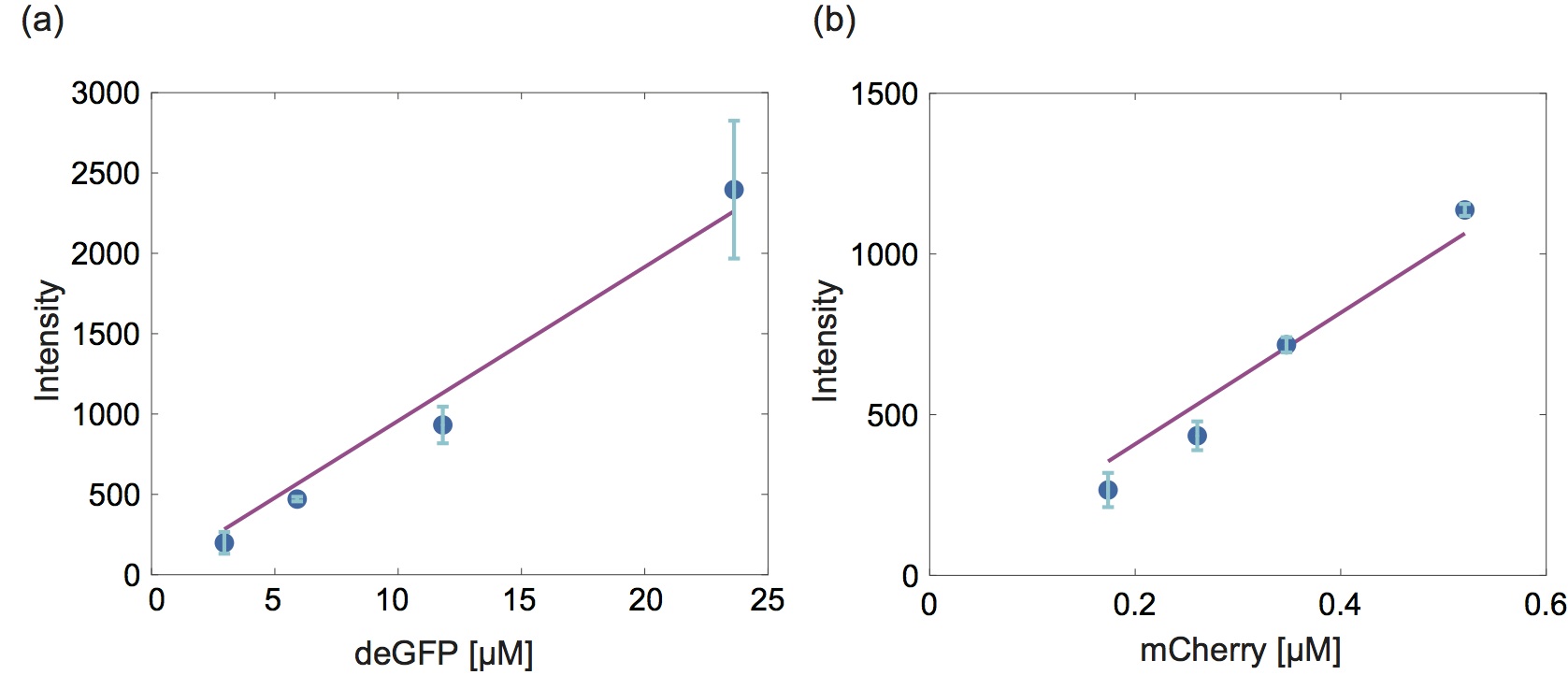}
\caption{\textbf{Calibration curves of the amount of marker proteins and their fluorescence intensity. }(a) Calibration curve of purified deGFP protein. (b) Calibration curve of purified mCherry protein. The error bars represent standard deviation. The solid curves are fitting curves.
}
\end{figure}

\subsection{The cell-free protein synthesis in concentrated TXTL reaction}

We tested the effect of increased volume fraction in the TXTL reaction in order to induce liquid-liquid phase separation.  The cell-free extract was concentrated by centrifugation to increase volume fraction until LLPS occurred. After that, plasmid DNA encoding mCherry gene was added and the expression level of synthesized mCherry protein was measured by using confocal fluorescent microscopy (Figure S2a). We found that the level of fluorescence of synthesized mCherry protein placed in the separated domain was only 1/10-fold of that in the TXTL reaction where LLPS was induced after cell-free gene expression (Figures S2a and S2b)). Although the expression of mCherry protein was detected at a higher level than the control (the absence of plasmid DNA), the activity of the TXTL reaction would be reduced by the concentration to induce LLPS (Figure S2c).

\begin{figure}[tb]
\centering
\includegraphics[scale=0.35,bb=0 0 992 792]{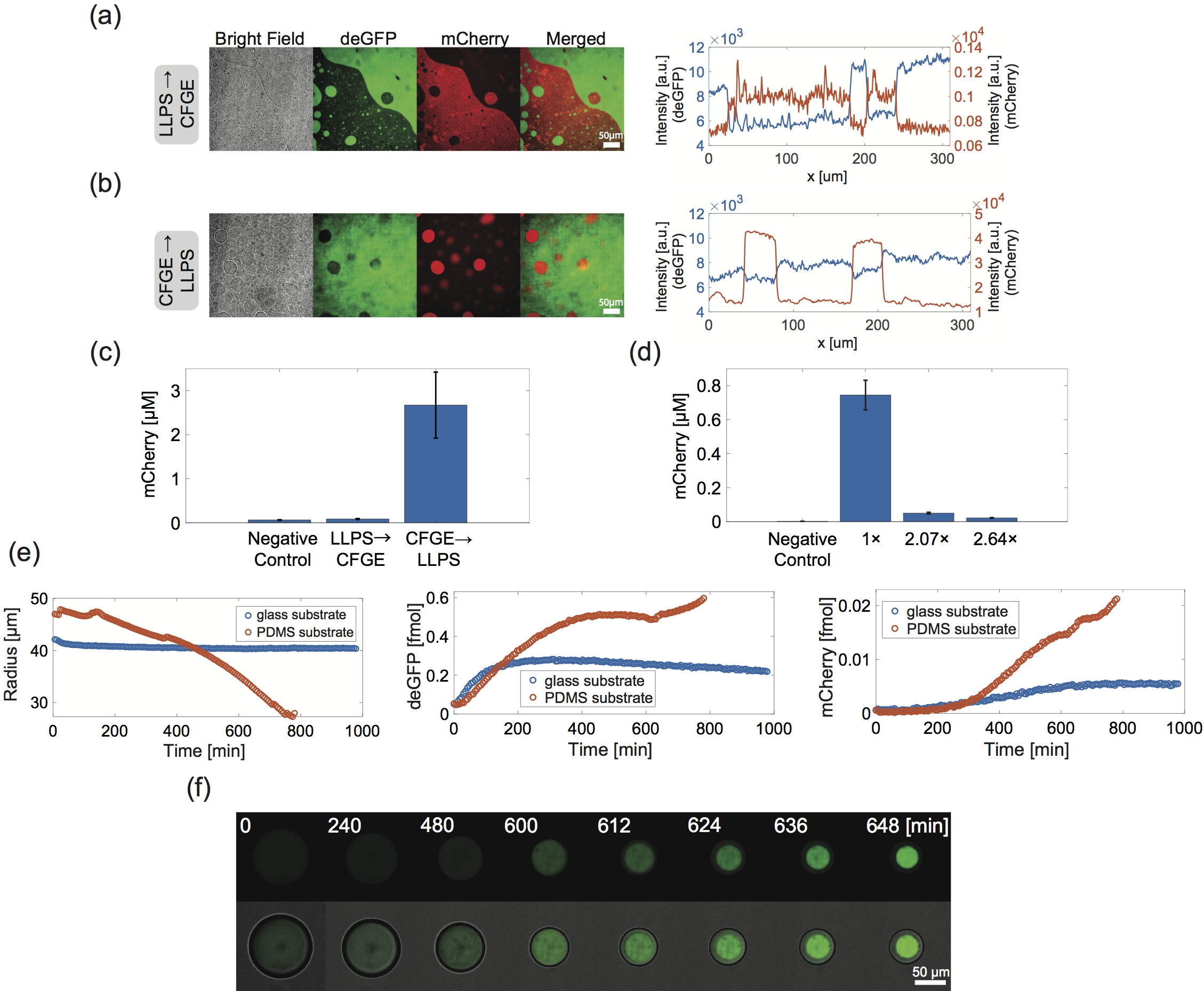}
\caption{\textbf{The activity of transcription-translation (TXTL) in the cell-free extract}. (a and b) The spatial distribution of purified GFP protein and synthesized mCherry protein expressed in the TXTL reaction for 10 hours. Left: confocal fluorescent images of segregated purified GFP protein (green) and synthesized mCherry protein (red) due to LLPS. Scale bar: \SI{50}{\micro\meter}. Right: Fluorescence of purified GFP protein. (a) Phase-separated domain was created with LLPS, and then cell-free gene expression was performed for \SI{10}{\hour}. This scheme is termed as a LLPS$\rightarrow$CFGE case. (b) The TXTL reaction was performed for \SI{10}{\hour} and then the extract containing the exrpressed mCherry was condensed for LLPS segregation. This scheme is termed as a CFGE$\rightarrow$LLPS case. (c) Total fluorescence of mCherry protein expressed under each condition. Control: the absence of plasmid DNA encoding mCherry gene, LLPS $\rightarrow$ CFGE case: mCherry gene was expressed after LLPS. CFGE$\rightarrow$LLPS case, mCherry gene was expressed before LLPS. The bar represents standard deviation. (d) Gene expression activity of the TXTL reaction using enriched extracts. The TXTL reaction solution was concentrated by dehydration using dialysis membranes, plasmid DNA (PLtetO1-mCherry) was expressed, and the gene expression activity was evaluated from the amount of mCherry synthesized. 1x, 2.07x, and 2.64x represent the concentration factor $\alpha_v$. The negative control represents the TXTL reaction without plasmid DNA. (e) The time evolution of droplet radius and cell-free gene expression in the droplet without evaporation effect. By replacing the PDMS substrate with the glass substrate, we can suppress the evaporation of water from the extract-in-oil droplet. (left) The shrinkage of droplet radius. Blue: the droplet onto the glass substrate (without evaporation), orange: the shrinking droplet onto the PDMS substrate. (center) Time course of synthesized deGFP protein. (right) Time course of synthesized mCherry protein. The blue plots (Glass substrate) represent the gene expression in the synthetic cell that does not show droplet shrinkage. (f) Time evolution of protein localization of purified deGFP marker protein in the TXTL reaction (2\% PEG) for the control experiment. Top: purified deGFP protein, Bottom: superimposed bright field image. Scale bar, \SI{50}{\micro\meter}.
}
\end{figure}

\subsection{The onset time of liquid-liquid phase separation}
To measure the time at which liquid-liquid phase separation (LLPS) starts, named as $T_{PS}$, we used the cross-correlation function of the fluorescence distributions of two fluorescent proteins, deGFP and mCherry. The cross-sectional fluorescent images of deGFP and mCherry taken by confocal microscopes are defined as $I_{G}(x,y,t)$ and $I_{R}(x,y,t)$, respectively. The cross-correlation function is defined by
\begin{equation}
C(t) = \frac{\sum_{x,y}(I_G(x,y,t) - \langle I_G \rangle_t)(I_R(x,y,t) - \langle I_R \rangle_t)}{\sqrt{(\sum_{x,y}(I_G(x,y,t) - \langle I_G \rangle_t)^2) (\sum_{x,y}(I_R(x,y,t) - \langle I_R \rangle_t)^2)}}
\end{equation}
where $\langle \cdot \rangle$ is the average over possible sites in the cross-section of the synthetic cell. Figure S3 shows this cross-correlation $C(t)$ calculated at each time. The time when $C(t)$ becomes the lowest value corresponds to the onset of LLPS of the TXTL reaction when the distributions of the two fluorescent proteins are exclusively separated. By obtaining this time for each synthetic cell, $T_{PS}$ at which LLPS starts is obtained.

\begin{figure}[tb]
\centering
\includegraphics[scale=0.59,bb=0 0 602 292]{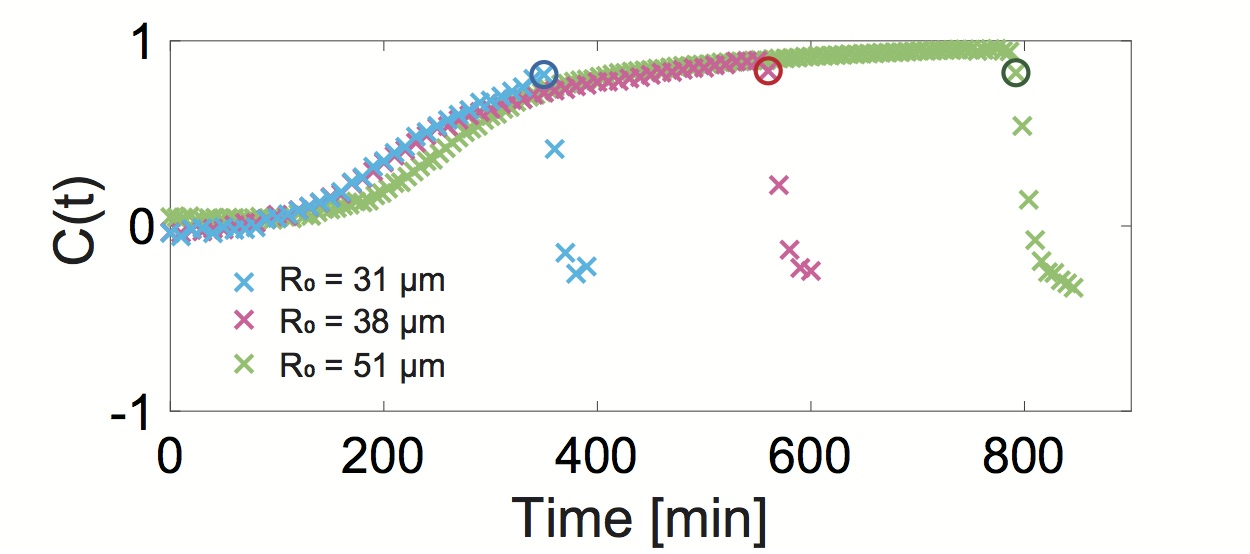}
\caption{\textbf{Time-course of cross correlation of spatial distribution of deGFP and mCherry inside the synthetic cells}. Black circle represents the time of LLPS, $T_{PS}$. We plotted the time course of correlation function obtained from three synthetic cells with different radius. Blue: $R_0 = \SI{31}{\micro\meter}$, Blue: $R_0 = \SI{38}{\micro\meter}$, Blue: $R_0 = \SI{51}{\micro\meter}$.
}
\end{figure}

\begin{figure}[h]
\centering
\includegraphics[scale=0.64,bb=0 0 552 292]{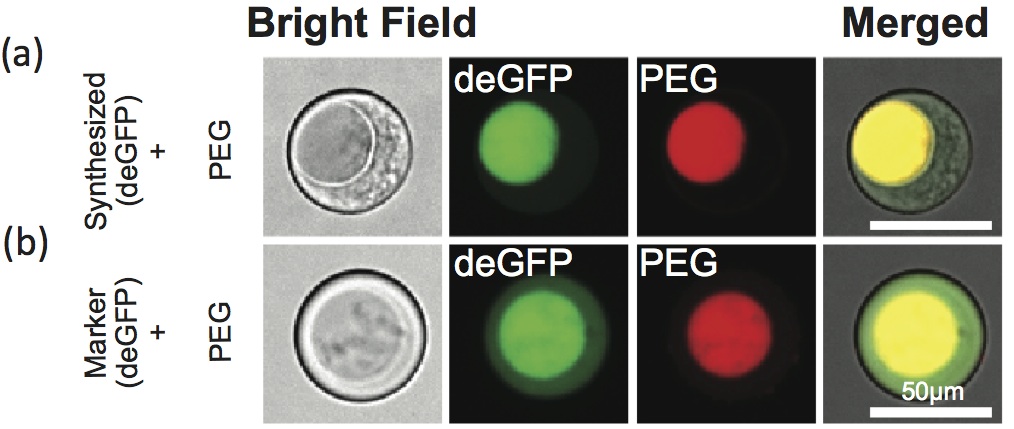}
\caption{\textbf{The confocal microscope images of PEG polymer and proteins after LLPS in the synthetic cells.} (a) The synthetic cell encapsulating Rhodamine-PEG and producing synthetic deGFP through the TXTL reaction. (b) The synthetic cell encapsulating Rhodamine-PEG and purified deGFP protein. (c) The synthetic cell encapsulating purified mCherry protein and purified deGFP protein. Scale bar: \SI{50}{\micro\meter}. 
}
\end{figure}

\begin{figure}[h]
\centering
\includegraphics[scale=0.59,bb=0 0 782 252]{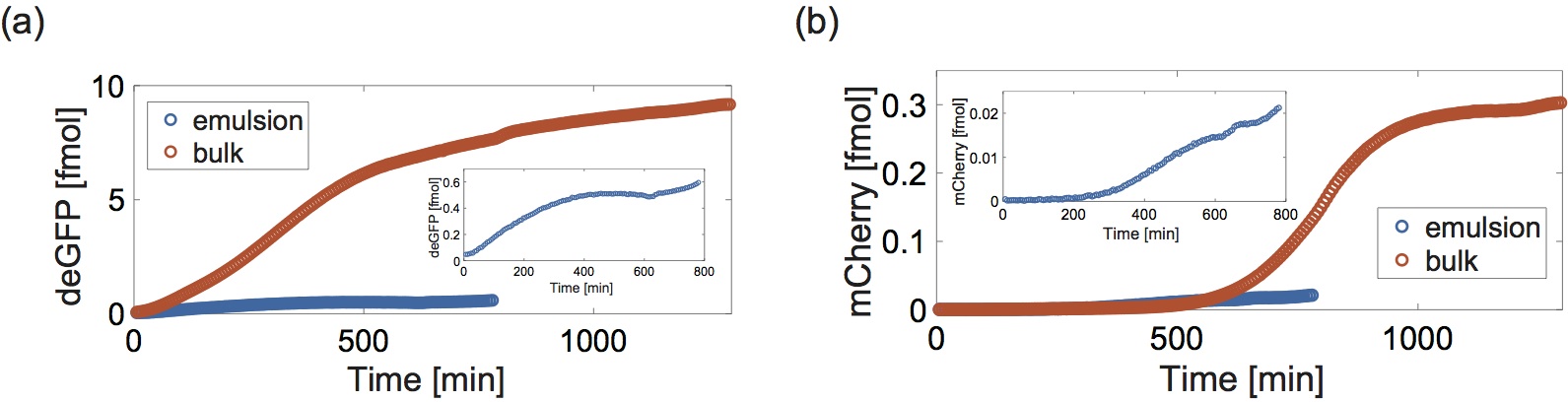}
\caption{Comparison of gene expression kinetics in the emulsion (blue, emulsion) and in the chamber without confined emulsion (orange, bulk). (a) Cell-free gene expression of deGFP. (b) Cell-free gene expression of mCherry.
}
\end{figure}

\subsection{Acceleration of LLPS formation in the synthetic cells}

\begin{figure}[h]
\centering
\includegraphics[scale=0.6,bb=0 0 392 292]{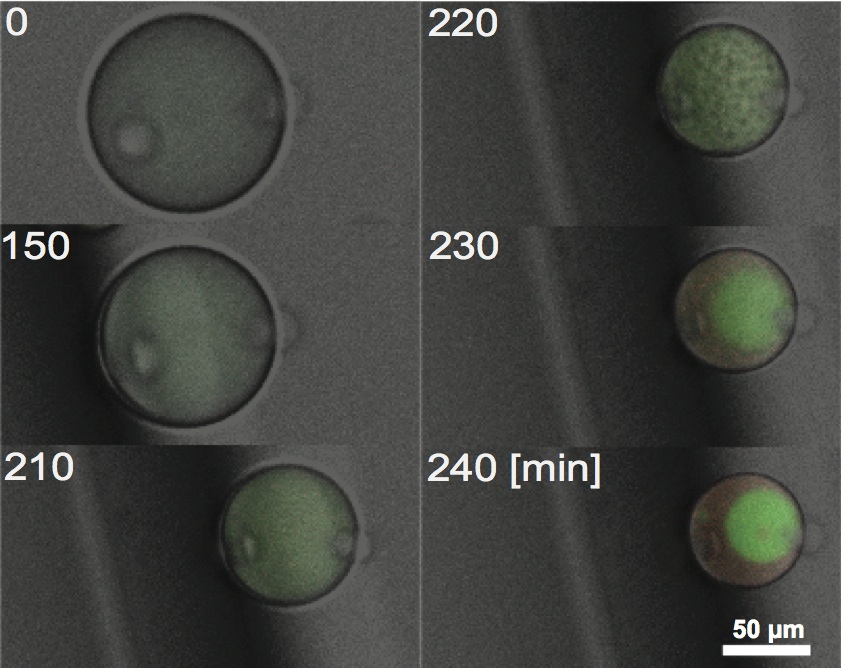}
\caption{\textbf{Acceleration of LLPS induced by the drift of the synthetic cell} (unit: min). When the oil-air interface moves the droplets in close proximity ($t = \SI{210}{\minute}$), the radius of the emulsion decreases rapidly and LLPS occurs at $t = \SI{230}{\minute}$. $R_0 = \SI{57}{\micro\meter}$. Scale bar, \SI{50}{\micro\meter}.
}\label{figs6}
\end{figure}

The onset of LLPS $T_{PS}$ took more than 10 h for large emulsions, e.g., $R_0 >\SI{50}{\micro\meter}$ because the time until phase separation started was determined by the rate of slow dehydration from the W/O interface. We examined if the compartmentalization in the TXTL reaction can be accelerated by changing the rate of droplet shrinkage under the flow of the external oil phase. Under the drifting flow at the oil-air interface layer, the emulsion with $R_0 = \SI{57}{\micro\meter}$ required only $T_{PS}$=230 min for phase separation owing to an increase in the rate of reduction in the radius (Figure S6). This shows that the LLPS onset time can be controlled using interfacial drift, and even large droplets can partition gene expression products within a few hours.

\subsection{Mathematical model of emulsion shrinkage}

The TXTL reaction for cell-free gene expression was encapsulated in a W/O droplet covered with a lipid monolayer. At the boundary of this droplet, water evaporated from the surface and absorbed into the oil phase in the form of small reverse micelles. Then, the volume of the droplet decreased by a constant amount per unit time, $\beta$.

The kinetics of droplet shrinkage was modeled as follows: The radius of the droplet is $R(t)$, and we assume that the volume of the droplet is $V(t)=\frac{4 \pi}{3} R^3(t)$ because it is spherical. As micelles are released from the surface of the droplet, the amount of reduction in water is proportional to the surface area of the droplet. The rate of volume reduction is calculated as

\begin{equation}
\frac{dV}{dt}= S\frac{dR}{dt}= - \beta S
\end{equation}
Hence, the reduction in droplet radius $R(t)$ is given by
\begin{equation}
\frac{dR}{dt}= - \beta 
\end{equation}
and 
\begin{equation}
R(t)= R_0 - \beta t
\end{equation}
where $R_0$ is the radius at the initial condition.

It was assumed that phase separation occurred when the reaction concentration reached a certain value. The initial concentration of the cell-free reaction was constant for all droplets. However, as the change in volume depended on the size, the time required for the onset of phase separation also depended on size. The concentration factor of the reaction at which phase separation begins is expressed as $\alpha_v = \frac{V_0}{V}$ ($\alpha_v > 1$ for shrinking emulsions). The volume of the droplet at the onset of phase separation, $V(T_{PS})=\frac{4 \pi}{3} R^3(T_{PS})$, is given by
\begin{equation}
V(T_{PS})=  \frac{V_0}{\alpha_v} = \frac{4 \pi}{3} \frac{R_0^3}{\alpha_v}
\end{equation}
Based on the volume of a spherical body, 
\begin{equation}
\frac{4 \pi}{3} (R_0 - \beta T_{PS})^3 =  \frac{4 \pi}{3} \frac{R_0^3}{\alpha_v},
\end{equation}
Finally, we obtain the time required for the onset of phase separation as
\begin{equation}
T_{PS} = \frac{1-(\alpha_v^{*})^{-\frac{1}{3}}}{\beta}R_0
\end{equation}

\begin{table}[tbh] \caption{\textbf{List of TXTL reaction solution, plasmid DNA and marker proteins used in this study.}}
\scalebox{0.8}{\begin{tabular}{|l|l|l|l|}
\hline 
\textbf{Figure} & \textbf{Reaction solution} & \textbf{Expression plasmids} & \textbf{Purified marker protein} \\ \hline
1(a), 1(b), 1(d) & TXTL(2\% PEG) & \multicolumn{1}{c|}{-} & deGFP: \SI{5.22}{\micro M} \\ \hline
1(c) & TXTL(2\% PEG) & \multicolumn{1}{c|}{-} & deGFP: \SI{5.22}{\micro M}, mCherry: \SI{4.34}{\micro M} \\ \hline
2(a)-2(g) & TXTL(2\% PEG) & PLtetO1-mCherry: 1.36 nM & deGFP: \SI{0.49}{\micro M} \\ \hline
2(h) & TXTL(2\% PEG) & PLtetO1-mCherry: 1.36 nM & deGFP: \SI{5.22}{\micro M} \\ \hline
3 & TXTL(2\% PEG) & PLtetO1-mCherry: 1.36 nM &\multicolumn{1}{c|}{-} \\ 
& & PLtetO1-deGFP: 1.39 nM & \\ \hline
4(a) & PBS(0\% PEG) & \multicolumn{1}{c|}{-} & deGFP: \SI{0.49}{\micro M}, mCherry: \SI{0.43}{\micro M} \\ \hline
4(b) & TXTL(2\% PEG) & \multicolumn{1}{c|}{-} & deGFP: \SI{0.49}{\micro M}, mCherry: \SI{0.43}{\micro M} \\ \hline
 4(c) & TXTL(0\% PEG) & \multicolumn{1}{c|}{-} & deGFP: \SI{0.49}{\micro M}, mCherry: \SI{0.43}{\micro M} \\ \hline
4(d) & PBS(2\% PEG) & \multicolumn{1}{c|}{-} & deGFP: \SI{0.49}{\micro M}, mCherry: \SI{0.43}{\micro M} \\ \hline
4(e) & TXTL(0 - 10\% PEG)& \multicolumn{1}{c|}{-} & \multicolumn{1}{c|}{-} \\ \hline
4(f): 2\% PEG & TXTL(2\% PEG) & PLtetO1-mCherry: 1.36 nM & deGFP: \SI{5.22}{\micro M} \\ \hline
4(f): 0\% PEG & TXTL(0\% PEG) & PLtetO1-mCherry: 1.36 nM & deGFP: \SI{5.22}{\micro M} \\ \hline
5 & TXTL(2\% PEG) & PLtetO1-mCherry: 0.78 nM & deGFP: \SI{5.22}{\micro M} \\ \hline
 \end{tabular} }
\end{table}

\end{document}